\documentclass[11pt,fleqn]{article}

\usepackage[left=2.5cm,
            right=2.5cm,
            top=2.5cm,
            bottom=3.5cm,
            foot=1.5cm]{geometry}
 
\usepackage{graphicx}
\usepackage{orcidlink}
\usepackage{alltt}
\usepackage{lmodern}
\usepackage{amssymb, amsfonts, amsthm, mathtools}
\usepackage{latexsym, hyperref, url, moreverb}
\usepackage{enumerate}
\usepackage{xspace}
\usepackage{booktabs}
\usepackage{xcolor}
\usepackage[utf8]{inputenc}
\usepackage[T1]{fontenc}

\usepackage[]{natbib}
\graphicspath{{Figures/}}

\makeatletter
\newcommand\code{\bgroup\@makeother\_\@makeother\~\@makeother\$\@codex}
\def\@codex#1{{\normalfont\ttfamily\hyphenchar\font=-1 #1}\egroup}
\makeatother

\IfFileExists{upquote.sty}{\usepackage{upquote}}{}

\renewcommand{\hat}[1]{\widehat{#1}}

\newcommand{\x}{\boldsymbol{x}}
\newcommand{\y}{\boldsymbol{y}}
\newcommand{\z}{\boldsymbol{z}}

\newcommand{\U}{\boldsymbol{U}}

\newcommand{\J}{\boldsymbol{J}}

\newcommand{\Deltab}{\boldsymbol{\Delta}}
\newcommand{\mub}{\boldsymbol{\mu}}

\newcommand{\Sigmab}{\boldsymbol{\Sigma}}

\newcommand{\Psib}{\boldsymbol{\Psi}}
\newcommand{\thetab}{\boldsymbol{\theta}}

\newcommand{\lambdab}{\boldsymbol{\lambda}}
\newcommand{\Lambdab}{\boldsymbol{\Lambda}}

\newcommand{\Space}{\mathcal{S}}
\newcommand{\XSpace}{\mathcal{X}}
\newcommand{\YSpace}{\mathcal{Y}}
\renewcommand{\hat}[1]{\widehat{#1}}

\DeclareMathOperator*{\argmax}{arg\,max}

\DeclareMathOperator{\Exp}{\mathbb{E}}

\DeclareMathOperator{\NCE}{\text{NCE}}

\usepackage{bbold}

\newcommand{\T}{{}^{\!\top}}

\newcommand{\Data}{\mathcal{D}}
\newcommand{\Xspace}{\mathcal{X}}

\newcommand{\Real}{\mathbb{R}}

\begin{document}

\title{\LARGE\bfseries%
A Model-Based Clustering Approach for Bounded Data Using Transformation-Based Gaussian Mixture Models}

\author{
\large      Luca Scrucca~\orcidlink{0000-0003-3826-0484}\\
\normalsize Department of Statistical Sciences, University of Bologna}

\date{\normalsize \today}

\maketitle

\begin{abstract}
\noindent%
The clustering of bounded data presents unique challenges in statistical analysis due to the constraints imposed on the data values. 
This paper introduces a novel method for model-based clustering specifically designed for bounded data. 
Building on the transformation-based approach to Gaussian mixture density estimation introduced by \cite{Scrucca:2019}, we extend this framework to develop a probabilistic clustering algorithm for data with bounded support that allows for accurate clustering while respecting the natural bounds of the variables.

In our proposal, a flexible range-power transformation is employed to map the data from its bounded domain to the unrestricted real space, hence enabling the estimation of Gaussian mixture models in the transformed space. 
Despite the close connection to density estimation, the behavior of this approach has not been previously investigated in the literature.
Furthermore, we introduce a novel measure of clustering uncertainty, the Normalized Classification Entropy (NCE), which provides a general and interpretable measure of classification uncertainty.
The performance of the proposed method is evaluated through real-world data applications involving both fully and partially bounded data, in both univariate and multivariate settings, showing improved cluster recovery and interpretability.

Overall, the empirical results demonstrate the effectiveness and advantages of our approach over traditional and advanced model-based clustering techniques that rely on distributions with bounded support.
\end{abstract}

\clearpage

\tableofcontents	

\clearpage

\section{Introduction}
\label{sec:intro}

Clustering is a fundamental task in data analysis, aiming to identify natural groupings or patterns within a dataset and thus facilitate the discovery of underlying structures and relationships among data points.
While numerous model-based clustering methods exist for unconstrained data, either having symmetric or asymmetric distributions, the clustering of bounded data presents unique challenges that require specialized approaches. 
Data with bounded support frequently arise in various fields, including economics, biology, environmental science, and social sciences. Examples include percentages, proportions, non-negative variables, e.g. arising from physical measurements, and any metric naturally limited to a fixed range.
Traditional clustering methods often struggle when applied to bounded data due to their inability to adequately account for the constraints imposed by the data bounds, leading to inaccurate inference and suboptimal clustering results. 

Model-based clustering, which assumes that data are generated from a mixture of probability distributions, provides a principled and flexible framework for clustering.
Gaussian mixture models (GMMs) assume the data are generated from a mixture of Gaussian distributions, with each component representing a distinct cluster. 
They are widely used due to their flexibility to model complex data structures and their adaptability to a variety of clustering problems.

However, the direct application of GMMs to bounded data is problematic due to the inherent constraints on data values. In this context, alternative model-based approaches that account for natural bounds in the data can be adopted by assuming bounded distributions for the component densities.

For instance, in the case of positive, right-skewed data, a mixture of gamma distributions \citep{John:1970} provides a natural alternative. \cite{Bagnato:Punzo:2013} and \cite{Young:etal:2019} proposed mixture of univariate unimodal gamma densities, while \cite{Wiper:Insua:Ruggeri:2001} discussed a similar model in a Bayesian framework, although also restricted to univariate data.
Another potential family of distributions is introduced in \cite{Karlis:Santourian:2009}, who employed a mixture of normal inverse-Gaussian distributions. This model extends naturally to the multivariate case, allowing for skewness, fat tails and, as by-product, can handle variables bounded at zero.

For data bounded at both the lower and upper limits, \cite{Bagnato:Punzo:2013} proposed the mixture of univariate unimodal beta densities. Building on this,  \cite{Dean:Nugent:2013} developed a model-based clustering procedure for data confined within the unit hypercube using a mixture of univariate unimodal beta distributions. However, their approach in the multivariate case is constrained by the assumption of conditional independence, whereby variables are considered conditionally independent given component membership.

A different approach was introduced in speech processing, where a bounded Gaussian mixture model \citep[BGMM;][]{Hedelin:Skoglund:2000} was employed and later extended to a bounded generalized Gaussian mixture model \citep[BGGMM;][]{Lindblom:Samuelsson:2003}. The BGGMM encompasses several models, including the standard GMM and BGMM, as special cases. These methods rely on truncated GMMs, where the unbounded Gaussian densities are multiplied by an indicator function that equal 1 if the component densities lie within the bounded data region and 0 otherwise; then, a subsequent normalization step ensures a proper marginal density distribution.

A common alternative to modeling bounded distributions directly is the use of data transformation techniques. Transforming bounded data onto an unbounded scale enables the application of standard statistical methods, then followed by an inverse transformation to express the results obtained in the original data space.
For instance, transformations such as the Box-Cox \citep{Lo:Gottardo:2012} and Manly \citep{Zhu:Melnykov:2018} transformations have been employed to manage skewness in component distributions.

More recently, \cite{Gallaugher:etal:2020} compared the performance of Gaussian mixtures in handling skewed data or outliers with that of mixtures of skewed distributions, such as the variance-gamma distribution \citep[VG;][]{McNicholas:etal:2017} and the generalized hyperbolic distribution \citep[GH;][]{Browne:McNicholas:2015}. They also examined mixtures incorporating transformations aimed at achieving near-normality, including the power transformation \citep{Yeo:Johnson:2000} and the Manly transformation \citep{Manly:1976}. Their results indicate that no single method consistently outperforms the others, with the optimal choice depending on the specific characteristics and context of the data being analyzed.

Most of the research on data transformation focuses on skewed data, where the main goal is to reduce skewness, making the data more symmetric and closer to a Gaussian distribution. However, if skewness varies across different clusters or subsets of the data, applying a single global transformation might not be appropriate.
In contrast, transformations for bounded data aim to convert data that is constrained within a specific range into an unbounded form, allowing for more flexible modeling. The underlying assumption is that a single transformation can be applied coordinate-wise to the entire dataset, as the bounding is consistent across all observations.

In this paper, we propose a novel model-based clustering method specifically designed for bounded data.
Our approach builds upon the range-power transformation framework for Gaussian mixtures proposed by \cite{Scrucca:2019}.
The method involves mapping bounded data into an unbounded space, where standard Gaussian mixture models can be applied, followed by an inverse transformation to recover the clustering results in the original bounded space. 
While the original framework was developed for density estimation, we extend it to the clustering setting. To the best of our knowledge, no previous work has directly addressed clustering within this framework, despite the close connection to density estimation.
We also introduce a novel measure of clustering uncertainty, called Normalized Classification Entropy (NCE), which provides a general and interpretable measure of classification uncertainty that can be applied to model-based clustering approaches more broadly.
The proposed approach has proven effective in real data application, offering a general and flexible framework for clustering bounded data.

In Section~\ref{sec:methods}, we first review the model-based approach to clustering, with particular emphasis on the Gaussian mixture model (GMM). 
The proposed approach is then presented, introducing the range-power transformation and its integration within the finite mixture model framework. 
We discuss maximum likelihood estimation via the EM algorithm, as well as model selection and methods for assessing clustering and classification uncertainty in this specific context.
Section~\ref{sec:applications} applies the method to real-data examples, covering both fully and partially bounded data in univariate and multivariate contexts.
We compare our method against both the standard GMM and more advanced model-based clustering techniques that incorporate distributions with bounded support specific to the analyzed variables.
Finally, Section~\ref{sec:conclusion} summarizes the main contributions of this work, discusses its strengths and limitations, and outlines potential directions for future research.

\clearpage

\section{Methods}
\label{sec:methods}

\subsection{Finite mixture modeling for clustering}
\label{sec:fmm}

Consider a multivariate dataset $\Data = \{ \x_i \}_{i=1}^n$ of $n$ observations, where each observation $\x_i$ is drawn from a $d$-dimensional random vector $\x$ with unbounded support $\Space_\Xspace \equiv \Real^d$. 

In model-based clustering we aim to partition the observations into $G$ distinct groups or clusters by typically employing a finite mixture model \citep[FMM;][]{McLachlan:Peel:2000, McLachlan:etal:2019}. 
Within this framework, each mixture component is directly associated with a cluster, effectively representing a distinct grouping of the data.
In its general form, a FMM can be expressed as:
\begin{equation}
f(\x; \Psib) = \sum_{k=1}^{G} \pi_k f_k(\x; \thetab_k),
\label{eqn:fmm}
\end{equation}
where $\pi_k$ represents the mixing proportions or weights, subject to the constraints $\pi_k > 0$ and $\sum_{k=1}^G \pi_k = 1$, and $f_k(\x; \thetab_k)$ denotes the multivariate density function of the $k$th component with parameters vector $\thetab_k$ ($k = 1, \dots, G$). 
If the density function $f_k()$ is usually assumed to be known, the parameters of the FMM, $\Psib = (\pi_1, \dots, \pi_{G-1}, \thetab_1, \dots, \thetab_G)$, are unknown and need to be estimated from the data.

One of the earlier, and still the most popular, model for continuous data is the Gaussian mixture model (GMM), which is obtained from \eqref{eqn:fmm} assuming a multivariate Gaussian distribution for each component density 
\citep{Fraley:Raftery:2002, Bouveyron:Celeux:Murphy:Raftery:2019, Gormley:Murphy:Raftery:2023, mclust:book:2023}.
A $G$-component GMM can be expressed as
\begin{equation}
f(\x; \Psib) = \sum_{k=1}^{G} \pi_k \phi(\x; \mub_k, \Sigmab_k),
\label{eqn:gmm}
\end{equation}
where $\phi(\x; \mub_k, \Sigmab_k)$ denotes the multivariate Gaussian density function with mean $\mub_k$ and covariance matrix $\Sigmab_k$ for the $k$-th mixture component. 

Parsimonious covariances decomposition can be obtained by imposing constraints on the 
geometric characteristics, such as volume, shape, and orientation, of corresponding ellipsoids. 
This is achieved using the following covariance matrices eigen-decomposition
\citep{Banfield:Raftery:1993, Celeux:Govaert:1995}:
\begin{equation}
\Sigmab_k = \lambda_k \U_k \Deltab_k \U\T_k,
\label{eqn:eigendecomp}
\end{equation}
where 
$\lambda_k = |\Sigmab_k|^{1/d}$ is a scalar controlling the volume, $\Deltab_k$ is a diagonal matrix controlling the shape, such that $|\Deltab_k| = 1$ and with the normalized eigenvalues of $\Sigmab_k$ in decreasing order, and $\U_k$ is an orthogonal matrix of eigenvectors of $\Sigmab_k$ that controls the orientation.
Further details on the resulting 14 different models can be found in \citet[Sec. 2.2.1]{mclust:book:2023}.

Maximum likelihood estimation of GMM parameters, $\Psib = (\pi_1, \dots, \pi_{G-1}, \mub_1, \dots, \mub_G, \Sigmab_1, \dots, \Sigmab_G)$, is commonly carried out using the Expectation-Maximization (EM) algorithm \citep{Dempster:Laird:Rubin:1977}. 
Let $\z_i$ denote the membership multinomial latent variable, with $z_{ik} = 1$ if observation $i$ belongs to cluster $k$, and $z_{ik} = 0$ otherwise. 
The EM algorithm iteratively maximizes the complete-data log-likelihood:
\begin{equation*}
\ell_C(\Psib) = \sum_{i=1}^n \sum_{k=1}^G z_{ik} \left\{ \log\pi_k + \log\phi(\x_i ; \mub_k,\Sigmab_k) \right\},
\end{equation*}
by alternating between two steps, the Expectation (E) step and the Maximization (M) step. 
For a comprehensive treatment of estimation using the EM algorithm and its properties, refer to \cite{McLachlan:Krishnan:2008}.
Details on the M-step for the different covariance parameterizations can be found in \cite{Celeux:Govaert:1995} and \cite[Sec. 2.2.2]{mclust:book:2023}.

\subsection{A transformation-based approach for Gaussian mixtures clustering}

A direct application of Gaussian mixtures, as formulated in \eqref{eqn:gmm}, may lead to inaccurate density estimates and erroneous clustering assignments when some or all the variables of the dataset are bounded.  
To overcome these limitations, we propose employing the transformation-based approach introduced by \cite{Scrucca:2019}. 
The core idea of this approach is to map the bounded variables to an unbounded space, where standard GMMs can be applied more effectively, and then transform the results back to the original bounded space, where clustering assignments can be straightforwardly obtained.
An additional advantage of this approach is that it enables the fitting of parsimonious models by using component-covariance eigen-decompositions from \eqref{eqn:eigendecomp}, which are particularly beneficial in high-dimensional settings for reducing complexity and improving the interpretability of the clustering results.

Let $\x$ represent a random vector from a distribution with bounded support $\Space_\XSpace \subset \Real^d$. 
Consider the family of continuous monotonic transformations, denoted by $\y = t(\x; \lambdab)$, where $\lambdab = [\lambda_1, \lambda_2, \dots, \lambda_d]\T \in \Lambdab$ represents the vector of transformation parameters, mapping the bounded support of data to an unbounded space, $\Space_\YSpace \equiv \Real^d$. 
Note that if one or more variables do not require transformation, the corresponding  $\lambda$ parameters can be set to 1 and kept fixed during the fitting process.

In the transformed space, the data can be modeled using a standard GMM:
\begin{equation*}
h(\y; \Psib) = \sum_{k=1}^{G} \pi_k \phi(\y; \mub_k, \Sigmab_k),
\end{equation*}
where $\Psib$ refers here to the parameters in the transformed scale, so $\mub_k$ and  $\Sigmab_k$ are, respectively, the mean vector and the covariance matrix of transformed variables in the $k$th component of the mixture.
Then, by applying the change-of-variable theorem, the density on the original scale can be recovered as:
\begin{equation}
f(\x; \Psib, \lambdab) = h(t(\x; \lambdab)) \times |\J(t(\x; \lambdab))|,
\label{eqn:dens}
\end{equation}
where $\J(t(\x; \lambdab))$ is the Jacobian of the transformation.
Since we consider coordinate-independent transformations
\begin{equation*}
t(\x; \lambdab) = [t(x_1; \lambda_1), t(x_2; \lambda_2), \dots, t(x_d; \lambda_d)]\T,
\end{equation*}
the Jacobian simplifies to a product of first derivatives:
\begin{equation}
\J(t(\x; \lambdab)) = 
\prod_{j=1}^d t'(x_j; \lambda_j).
\label{eqn:jacobian}
\end{equation}

The transformation $\y = t(\x; \lambdab)$ adopted is based on the \emph{range-power transformation} introduced in \cite{Scrucca:2019}, which is suitable for both partially and completely bounded data. 
For a variable $x_j \in (l_j, +\infty)$, where $l_j > -\infty$ represents its lower bound, the range-power transformation is defined as:
\begin{equation}
t(x_j; \lambda_j) = 
\begin{cases}
\dfrac{(x_j - l_j)^{\lambda_j} - 1}{\lambda_j} & 
\quad\text{if}\; \lambda_j \ne 0 \\[2ex]
\log(x_j - l_j) & 
\quad\text{if}\; \lambda_j = 0,
\end{cases}
\label{eqn:range_power_transform_lb}
\end{equation}
with continuous first derivative equal to 
\begin{equation}
t'(x_j; \lambda_j) = \frac{\partial t(x_j; \lambda_j)}{\partial x_j} = (x_j - l_j)^{\lambda_j-1}
\label{eqn:range_power_transform_lb_deriv}
\end{equation}
for any $\lambda_j \in \Lambda$ ($j= 1, \dots, d$).

If a variable $x_j \in (l_j, u_j)$, where $l_j > -\infty$ and $u_j < +\infty$ represent the lower and upper bounds, respectively, the range-power transformation is defined as:
\begin{equation}
t(x_j; \lambda_j) = 
\begin{cases}
\dfrac{ \left( \dfrac{x_j - l_j}{u_j - x_j} \right)^{\lambda_j} - 1}{\lambda_j}
  & \quad\text{if}\; \lambda_j \ne 0 \\[3ex]
\log \left( \dfrac{x_j - l_j}{u_j - x_j} \right)
  & \quad\text{if}\; \lambda_j = 0,
\end{cases}
\label{eqn:range_power_transform_lb_ub}
\end{equation}
with continuous first derivative given by
\begin{equation}
t'(x_j; \lambda_j) = \frac{\partial t(x_j; \lambda_j)}{\partial x_j} =
\begin{cases}
\left( \dfrac{x_j - l_j}{u_j - x_j} \right)^{\lambda_j-1} \dfrac{u_j - l_j}{(u_j -x_j)^2} 
  & \text{if}\; \lambda_j \ne 0 \\[3ex]
\dfrac{1}{x_j - l_j} + \dfrac{1}{u_j - x_j}
  & \text{if}\; \lambda_j = 0.
\end{cases}
\label{eqn:range_power_transform_lb_ub_deriv}
\end{equation}

Equations \eqref{eqn:range_power_transform_lb_deriv} and  \eqref{eqn:range_power_transform_lb_ub_deriv} can be used to compute the Jacobian in \eqref{eqn:jacobian}. Together with the range-power transformations in equations \eqref{eqn:range_power_transform_lb} and \eqref{eqn:range_power_transform_lb_ub}), these allow for the estimation of the mixture density in \eqref{eqn:dens}. 
However, both the unknown mixture parameters $\Psib$ and the transformation parameters $\lambdab$ need to be estimated, as discussed in Section~\ref{sec:mle}.

The range-power transformation was selected for its flexibility in accommodating both lower- and upper-bounded data while preserving interpretability. Alternative families of transformations could also be considered. For instance, the \citet{Manly:1976} transformation can handle both positive and negative values, although it does not naturally account for bounded support. The \citet{Yeo:Johnson:2000} transformation extends the Box-Cox approach to allow for both positive and negative values, but it similarly does not accommodate data with explicit bounds. In contrast, the range-power transformation is specifically designed for bounded data, making it particularly well-suited for the settings considered in this work.

\begin{figure}[htb]
\centering
\includegraphics[width=0.49\textwidth]{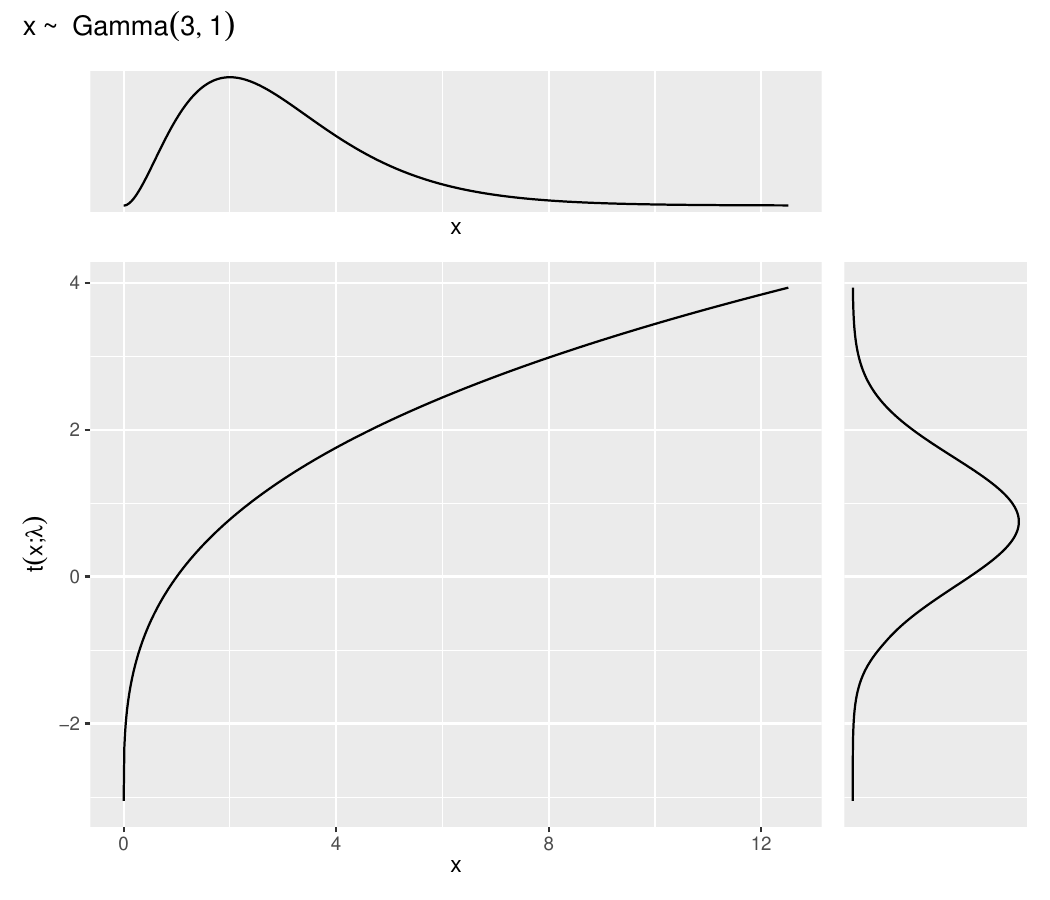}
\includegraphics[width=0.49\textwidth]{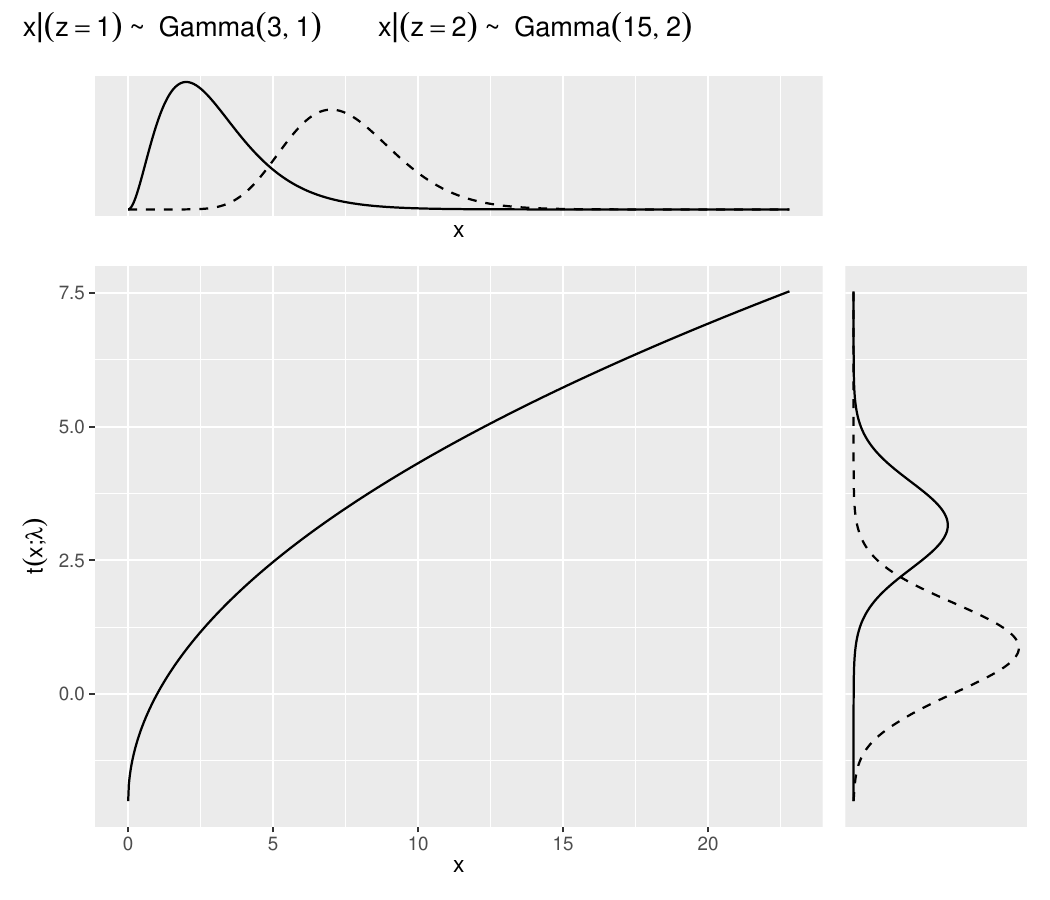}\\
\includegraphics[width=0.49\textwidth]{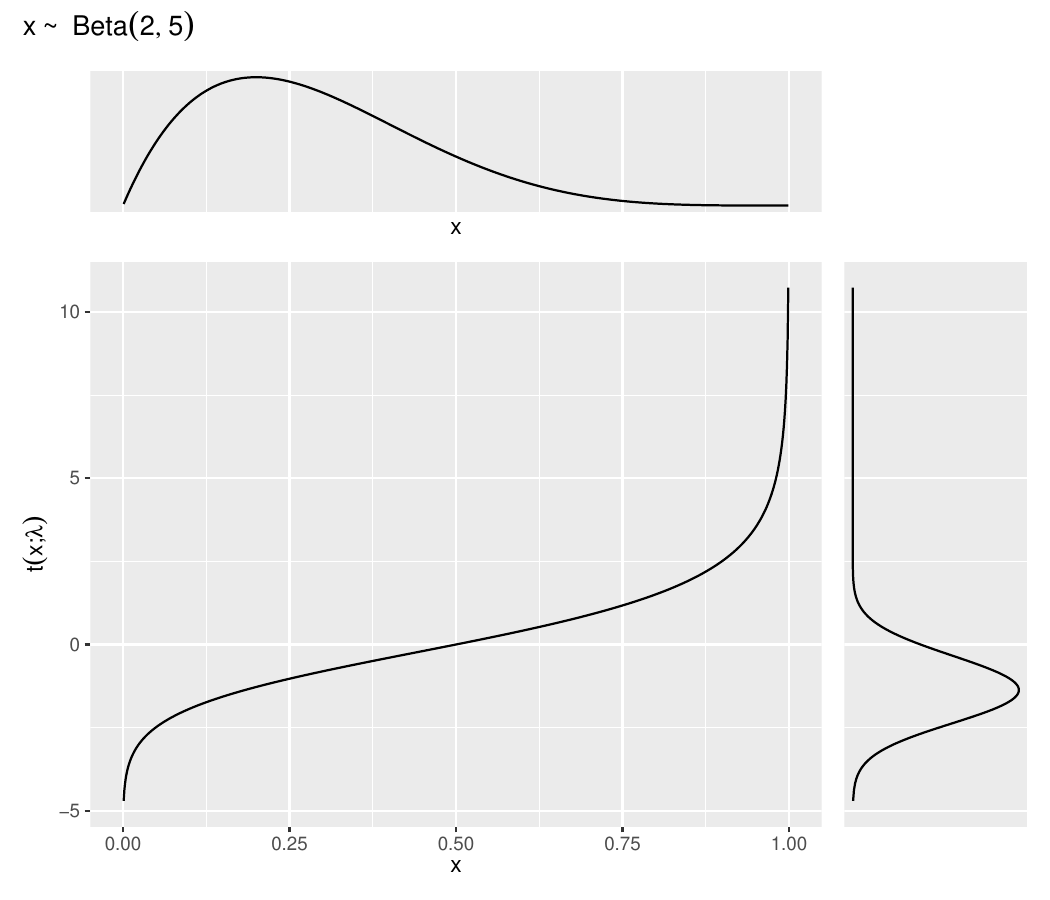}
\includegraphics[width=0.49\textwidth]{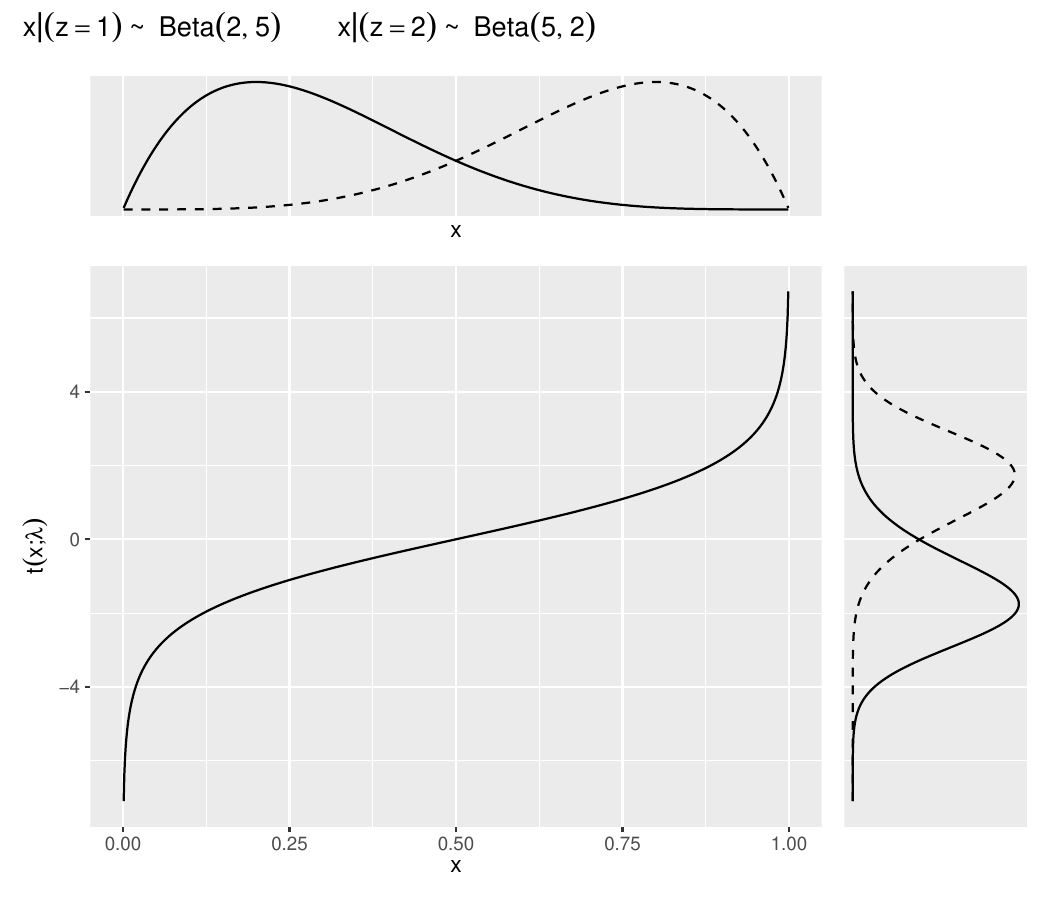}
\caption{Plots of range-power transformations for single-component distributions (left panels) and two-component mixtures (right panels) for univariate bounded data. The top panels show data with only a lower bound, while the bottom panels show data with both lower and upper bounds. Each panel displays the curve of the optimal transformation and the corresponding marginal densities on both the original and transformed scales.}
\label{fig:plots_trans}
\end{figure}

Figure~\ref{fig:plots_trans} illustrates the behavior of the range-power transformation on selected single- and two-component mixtures for univariate bounded data. 
Each panel shows the curve of the optimal transformation along with the corresponding marginal densities on both the original and transformed scales. 
The left panels contain the transformations for single-component distributions. In particular, the top-left panel presents the transformation for a lower-bounded Gamma$(3, 1)$ distribution, which results in a smooth, monotonic transformation. The bottom-left panel shows the transformation for a Beta$(2, 5)$ distribution, which is bounded on both sides. The transformation is nonlinear, with stronger effects near the boundaries due to the bounded support.  
The right panels display the transformations for two-component mixtures of bounded distributions. The top-right panel shows the transformation for a mixture of two Gamma distributions, namely Gamma$(3, 1)$ and Gamma$(15, 2)$. The bimodal structure of the mixture is evident from the density plot, and the transformation effectively reflects the underlying mixture components. The bottom-right panel illustrates the transformation for a mixture of Beta$(2, 5)$ and Beta$(5, 2)$ distributions. The bimodal nature is clearly visible, and the transformation adapts to the complexity of the mixture structure.  
These examples demonstrate the flexibility of the range-power transformation approach in handling different types of bounded data, including single and mixture distributions, while preserving the main clustering structure, if any, of the data.

\subsection{Maximum likelihood estimation via the EM algorithm}
\label{sec:mle}

Recalling the density in \eqref{eqn:dens}, the log-likelihood of the observed data can be expressed as:
\begin{equation*}
\ell(\Psib, \lambdab) = \sum_{i=1}^{n} \log \left( \sum_{k=1}^{G} \pi_k \phi(t(\x_i; \lambdab); \mub_k, \Sigmab_k) \times |\J(t(\x_i; \lambdab))| \right).
\end{equation*}
Maximum likelihood estimation can be pursued via the EM algorithm by maximizing the complete-data log-likelihood:
\begin{equation*}
\ell_C(\Psib, \lambdab) = \sum_{i=1}^n \sum_{k=1}^G z_{ik} \left\{ \log\pi_k + \log\phi(t(\x_i; \lambdab); \mub_k, \Sigmab_k) + \log |\J(t(\x_i; \lambdab))| \right\},
\end{equation*}
where $\z_i$ is the latent variable for cluster membership described in Section~\ref{sec:fmm}. 

At iteration $m$ of the EM algorithm, the conditional expectation of the complete-data log-likelihood given the observed data and the current parameter values can be expressed as:
\begin{equation*}
Q(\Psib,\lambdab; \Psib^{(m)},\lambdab^{(m)}) = 
  \sum_{i=1}^n \sum_{k=1}^G \hat{z}_{ik}^{(m+1)} 
  \left\{ \log\pi_k + 
          \log\phi(t(\x_i; \lambdab); \mub_k, \Sigmab_k) +
          \log|\J(t(\x_i; \lambdab))|
  \right\},
\end{equation*}
where $\hat{z}_{ik}^{(m+1)} = \Exp(I(\z_i=k)|t(\x_i,\hat{\lambdab}^{(m)}), \Psib^{(m)})$. 
Therefore, in the E-step the posterior probabilities are updated using
\begin{equation*}
\hat{z}_{ik}^{(m+1)} = 
  \dfrac{\hat{\pi}_k^{(m)} 
         \phi\left(t(\x_i; \hat{\lambdab}^{(m)}); 
                   \hat{\mub}_k^{(m)}, \hat{\Sigmab}_k^{(m)}
             \right)}
        {\displaystyle
         \sum_{g=1}^G \hat{\pi}_g^{(m)} 
         \phi\left(t(\x_i; \hat{\lambdab}^{(m)});
                   \hat{\mub}_g^{(m)}, \hat{\Sigmab}_g^{(m)}
             \right)}.
\end{equation*} 

In the M-step, the parameters $(\Psib, \lambdab)$ are updated by maximizing the $Q$-function, given the previous values of the parameters and the updated posterior probabilities. 
Following the Expectation-Conditional-Maximization (ECM) algorithm introduced by \cite{Meng:Rubin:1993}, this maximization is carried out in two steps.
In the first step, the updated value $\hat{\lambdab}^{(m+1)}$ is computed by maximizing the Q-function with respect to $\lambdab$. Since no closed-form solution is available, this requires numerical optimization, such as a Newton-type algorithm or a related method. 
Our implementation uses the \mbox{L-BFGS-B} method \citep{Byrd:etal:1995} available in the \code{optim()} function for the \textsf{R} statistical software \citep{Rstat}.
The remaining parameters are then obtained as in standard EM algorithm but accounting for the updated transformation parameters $\hat{\lambdab}^{(m+1)}$, i.e.
\begin{equation*}
\hat{\pi}_k^{(m+1)} = \dfrac{\sum_{i=1}^n \hat{z}_{ik}^{(m+1)}}{n}
\qquad\text{and}\qquad
\hat{\mub}_k^{(m+1)} = 
  \dfrac{\sum_{i=1}^n \hat{z}_{ik}^{(m+1)} t(\x_i; \hat{\lambdab}^{(m+1)})}
        {\sum_{i=1}^n \hat{z}_{ik}^{(m+1)}}.
\end{equation*}
The update formula for the covariance matrices depends on the assumed eigen-decomposition model. In the most general case of unconstrained covariance matrices, i.e.\ the VVV model in \texttt{mclust} nomenclature \citep[see][Table 2.1]{mclust:book:2023}, we have 
\begin{equation*}
\hat{\Sigmab}_k^{(m+1)} = 
  \dfrac{\displaystyle \sum_{i=1}^n \hat{z}_{ik}^{(m+1)} 
         \left(t(\x_i; \hat{\lambdab}^{(m+1)}) - \hat{\mub}_{k}^{(m+1)}\right)
         \left(t(\x_i; \hat{\lambdab}^{(m+1)}) - \hat{\mub}_{k}^{(m+1)}\right)\T}
        {\displaystyle \sum_{i=1}^n \hat{z}_{ik}^{(m+1)}}.
\end{equation*}

The initialization of the EM algorithm introduced above requires initial estimates of optimal marginal transformation parameters and an initial partition of the data.
The former can be obtained by numerically maximizing the marginal log-likelihood (using the \mbox{L-BFGS-B} method as mentioned above), i.e. by estimating
$\lambda_j^{(0)} = \argmax_{\lambda_j} \ell(\lambda_j)$, where
\begin{equation*}
\ell(\lambda_j) = \sum_{i=1}^{n} \log \phi(t(x_{ij}; \lambda_j); m_j, s^2_j) + \log t'(x_{ij}; \lambda_j),
\end{equation*}
with $x_{ij}$ being the $i$th observation on variable $j$, so $t(x_{ij}; \lambda_j)$ is the corresponding range-power transformation for parameter $\lambda_j$, with mean $m_j$ and variance $s^2_j$ ($j=1,\dots,d$).
The initial partition can then be obtained using the final classification from a $k$-means algorithm on the range-power transformed variables.
Alternatively, partitions obtained from model-based agglomerative hierarchical clustering \citep[MBAHC;][]{Scrucca:Raftery:2015} can be used. 
This initial partition of data points is used to start the algorithm from the M-step. 
Finally, the EM algorithm is stopped when the log-likelihood improvement falls below a specified tolerance value or a maximum number of iterations is reached.

Model selection in finite mixture modeling is typically carried out using information criteria, such as the Bayesian Information Criterion \citep[BIC;][]{Schwarz:1978}, or the Integrated Complete-data Likelihood criterion \citep[ICL;][]{Biernacki:Celeux:Govaert:2000}.
Both criteria penalize model complexity, favoring more parsimonious models unless the additional parameters are justified by a significant improvement in the likelihood. Additionally, ICL introduces an extra penalization for clustering overlap, promoting models that produce well-separated clusters.

Lastly, we note that estimating the $\lambda$ parameters introduces additional computational overhead to the fitting procedure, primarily due to the increased complexity of the likelihood surface rather than the number of parameters alone. 
The number of EM iterations required for convergence — and its impact on runtime — depends on problem-specific factors, as EM-based optimization is often more sensitive to the curvature of the parameter space than to its dimensionality.
Empirical results suggest that while estimating $\lambda$ parameters increases runtime, the additional cost remains reasonable in practical scenarios. For instance, on the wholesale dataset discussed in Section~\ref{sec:wholesale}, the median runtime increased by a factor of approximately four (89 ms vs. 22 ms when $\lambda$ parameters were estimated rather than fixed).

\subsection{Clustering and classification uncertainty}

In model-based clustering, assigning a data point to one of the identified clusters is straightforward using the maximum a posteriori (MAP) principle.
According to MAP, each observation is assigned to the cluster that has the highest posterior probability.

Consider a partition of the data $\Data = \{ \x_i \}_{i=1}^n$ into $G$ clusters, denoted as $\mathcal{C} = \{ C_1, C_2, \dots, C_G \}$, where $C_k \,\cap\, C_g = \emptyset$ (for $k \ne g$) and $\bigcup_{k=1}^{G} C_k = \Data$. 
The MAP procedure assigns an observation $\x_i$ to a cluster $C_{\hat{k}}$ according to the rule:
\begin{equation*}
\x_i \in C_{\hat{k}} 
\qquad\text{with}\quad 
\hat{k} = \argmax_{k \in \{1, \dots, G\}}\; \hat{z}_{ik},
\end{equation*}
where $\hat{z}_{ik}$ represents the posterior probability of observation $\x_i$ belonging to cluster $k$, i.e. 
\begin{equation*}
\hat{z}_{ik} = 
  \dfrac{\hat{\pi}_k 
         \phi\left(t(\x_i; \hat{\lambdab}); 
                   \hat{\mub}_k, \hat{\Sigmab}_k
             \right)}
        {\sum_{g=1}^G \hat{\pi}_g 
         \phi\left(t(\x_i; \hat{\lambdab});
                   \hat{\mub}_g, \hat{\Sigmab}_g
             \right)}.
\end{equation*} 

Hard classification based on the MAP principle assumes the most likely cluster is the correct one. However, once each data point is assigned to its most probable cluster, it becomes essential to evaluate the uncertainty of these assignments. 
Assessing clustering uncertainty is thus crucial, as some data points may not clearly belong to a single cluster, especially when they lie near cluster boundaries. This also allows to distinguish between well-separated clusters and those that exhibit substantial overlap.

To measure this uncertainty, we can examine the distribution of posterior probabilities across all clusters rather than focusing solely on the maximum probability. The uncertainty for each data point is captured by the following index:
\begin{equation*}
u_i = 1 - \max_{k} \hat{z}_{ik}.
\end{equation*}
This score ranges from 0 to $(G-1)/G$, with values close to 0 indicating low classification uncertainty, while values near the upper bound suggest greater uncertainty.

An alternative measure of classification uncertainty can be derived from the entropy. For each data point, we can compute 
\begin{equation*}
e_i = - \sum_{k=1}^G \hat{z}_{ik} \log(\hat{z}_{ik}),
\end{equation*}
which yields values in the range $[0,\log(G)]$, with higher entropy values indicating greater uncertainty in the assignment.
For easier interpretation, a normalized version of this classification entropy, ranging between 0 and 1, can be calculated by taking the ratio of each value to its maximum, i.e.%
\begin{equation}
e_i^* = \frac{e_i}{\log(G)} \in [0,1].
\label{eq:nclent}
\end{equation}%
An overall measure of classification uncertainty can thus be obtained by averaging the normalized entropy values in \eqref{eq:nclent} across all data points, yielding the \textit{Normalized Classification Entropy} (NCE):
\begin{equation}
\NCE = \frac{1}{n} \sum_{i=1}^n e_i^*
     = \frac{\sum_{i=1}^n  e_i}{n \log(G)},
\label{eq:nce}
\end{equation}%
for $G > 1$, and with the implicit assumption that $\NCE = 0$ when $G = 1$.
The index $\NCE$ quantifies the clustering uncertainty, taking a value of zero when each data point is assigned to its respective cluster with probability 1. 
As uncertainty increases, the index rises accordingly, reaching its maximum value of one when $z_{ik} = 1/G$ for all observations $i=1,\dots,n$, and clusters $k=1,\dots,G$.

The above proposal is reminiscent of the \textit{Normalized Entropy Criterion} (NEC) introduced by \citet{Celeux:Soromenho:1996}, and later improved by \citet{Biernacki:Celeux:Govaert:1999}, which also uses the classification entropy $E=\sum_{i=1}^n  e_i$ as in the numerator of equation~\eqref{eq:nce}, but applies a different denominator for scaling.
Moreover, the goals of NEC and NCE are fundamentally different. NEC was proposed as a criterion for model selection, whereas our proposal aims to measure the uncertainty in the clustering partition obtained.
For this reason, we caution against interpreting a very small NCE as necessarily indicative of a better clustering result. A low NCE suggests well-separated clusters with low uncertainty in the MAP assignments. On the contrary, an increase in NCE may reflect substantial overlap among clusters. While this could indicate poor clustering, it often reflects the inherent nature of the data, where the clusters are not well separated. Therefore, while NCE provides a useful measure of overall clustering uncertainty, it should not be used as the sole criterion for evaluating the quality of a clustering solution.

\clearpage

\section{Applications}
\label{sec:applications}

\subsection{Enzyme data}
\label{sec:enzyme}

The enzyme data set, originally described by \cite{Bechtel:etal:1993}, contains enzymatic activity measurements in the blood for an enzyme involved in the metabolism of carcinogenic substances. 
Enzymatic activity is usually greater than zero, as some baseline activity may be observed even in healthy individuals. However, such activity can be recorded as zero when the enzyme is completely absent, fully inhibited, or even if some minimal activity is present but falls below the detection limit of a laboratory test.
The analyzed data were collected from a sample of $n = 245$ unrelated individuals. Here, the primary interest lies in identifying subgroups of slow or fast metabolizers, which serve as markers of genetic polymorphism in the general population.

This benchmark data set for mixture models is notable for containing at least two components, one of which exhibits clear skewness. 
\cite{Richardson:Green:1997} employed a Bayesian GMM estimated using reversible jump MCMC to analyze the distribution of enzymatic activity. Their analysis suggests that a model with 3 to 5 mixture components is plausible, with the three-component solution being preferred, particularly in line with a simple underlying genetic model. In contrast, \cite{Karlis:Santourian:2009} applied a Normal Inverse Gaussian Mixture (NIGM) and identified two distinct, non-overlapping clusters. 

\begin{figure}[htb]
\centering
\includegraphics[width=\textwidth]{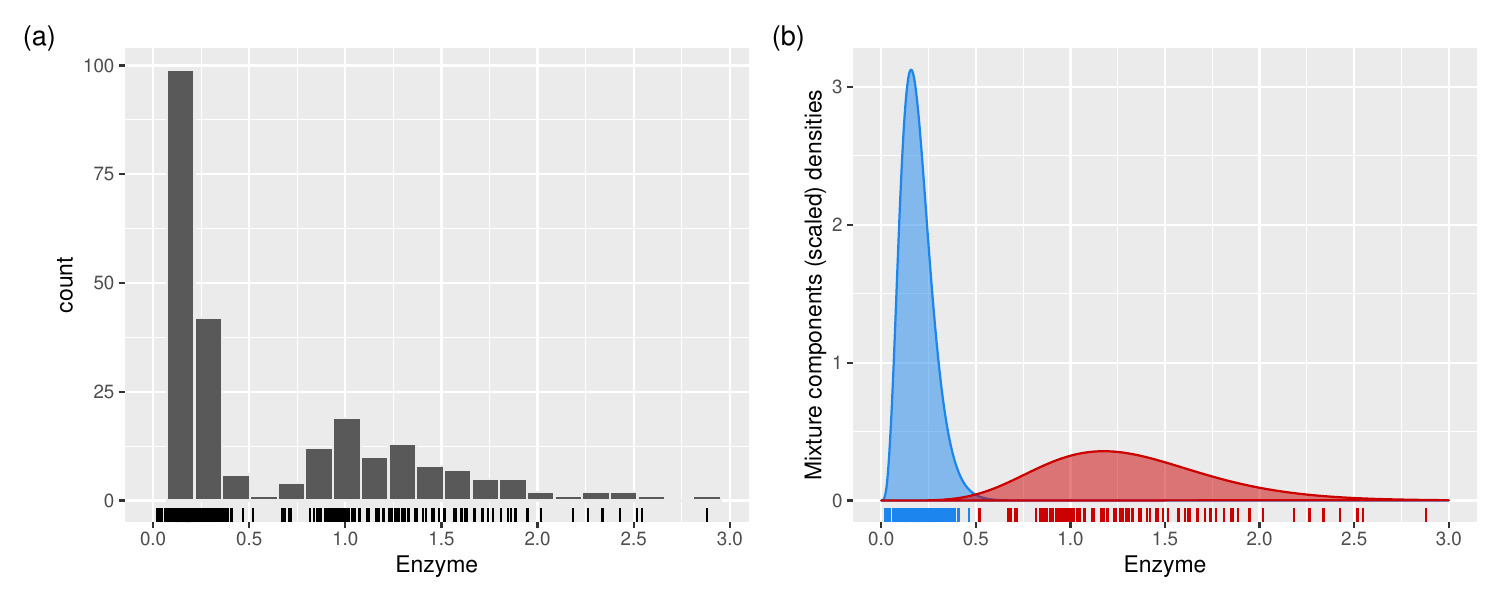}
\caption{(a) Histogram of the enzyme data and (b) plot of component densities, scaled by the corresponding mixing probabilities, estimated using the Gaussian Mixture Model for Bounded data (GMMB).}
\label{fig:enzyme1}
\end{figure}

Figure~\ref{fig:enzyme1}a presents the histogram of the enzyme data, while Figure~\ref{fig:enzyme1}b shows the estimated densities from the selected two-component GMMB model with unconstrained variances and a range-power transformation parameter in \eqref{eqn:range_power_transform_lb} equals to $\hat{\lambda} = 0.3666$.

This model, identified as the best fit according to both the BIC and ICL criteria, distinguishes between two clusters: one consisting of 152 observations characterized by a skewed distribution and enzymatic activity levels below 0.5, and a second, comprising 93 observations, forming a nearly symmetric group with enzymatic activity levels above 0.5. 
These results align with those reported by \cite{Karlis:Santourian:2009} and have been reproduced here using the function \code{MGHM()} from the \textbf{MixtureMissing} R package \citep{Rpkg:MixtureMissing}. 

\begin{figure}[htb]
\centering
\includegraphics[width=0.6\textwidth]{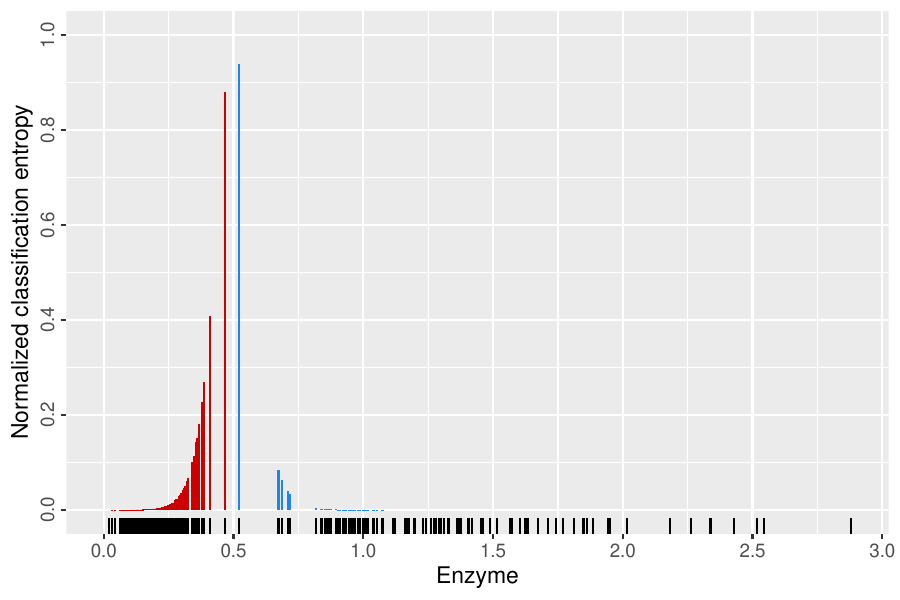}
\caption{Normalized classification entropy for each data point in the enzyme dataset showing clustering uncertainty.}
\label{fig:enzyme2}
\end{figure}

Although the GMMB model provides a better fit, as indicated by higher BIC and ICL values, and is more parsimonious with fewer parameters to estimate, it reflects a slightly larger classification uncertainty than the NIGM model, as measured by NCE. 
Figure~\ref{fig:enzyme2} shows the normalized classification entropy values from \eqref{eq:nclent}, which represent the contribution of each data point to the overall classification uncertainty provided by NCE. As can be clearly seen, most of the contribution comes from those observations near the classification boundary at an enzymatic level of $0.5$.
In any case, both models represent a clear improvement over the standard GMM, as shown in Table~\ref{tbl:enzyme}.

\begin{table*}[htb]
\caption{Model comparison for the clustering of the enzyme data using Gaussian Mixture Model (GMM), mixture of Normal Inverse Gaussian (NIGM) distributions, and Gaussian Mixture Model for Bounded data (GMMB), with the number of components selected according to the BIC criterion.
The nomenclature (V,2) for both the GMM and GMMB models denotes two-component heteroscedastic mixture models (i.e. models with different variances).}
\label{tbl:enzyme}
\centering
\begin{tabular}{@{}lrrrrr@{}}
\toprule
Model      & log-likelihood & df & BIC & ICL & NCE \\
\midrule
GMM(V,2)   & -54.6401 & 5 & -136.7865 & -148.9526 & 0.1109 \\
NIGM(2)    & -41.4723 & 9 & -132.4558 & -132.6530 & 0.0052 \\
GMMB(V,2)  & -46.1870 & 6 & -125.3815 & -127.9385 & 0.0208 \\
\bottomrule
\end{tabular}
\end{table*}

\clearpage

\subsection{Wholesale customer segmentation}
\label{sec:wholesale}

Customer segmentation aims to divide customers into distinct groups based on common characteristics, such as purchasing behavior. This partition helps businesses to tailor their marketing efforts, products, and services to specific customer segments, potentially maximizing customer satisfaction and profitability. 

The wholesale customers dataset, freely available on the UCI Machine Learning Repository \citep{uci_wholesale}, 
is a widely-used dataset to explore clustering and segmentation techniques. 
The dataset includes 440 observations, each representing a customer of a wholesale distributor. For each client, annual spending, in monetary units, is recorded for six product categories: fresh, milk, grocery, frozen, detergents paper, and delicatessen. 
Two additional categorical variables are available, corresponding to the region and customer channel.
Note that, unlike other authors who have analyzed this dataset, such as \cite{Punzo:Tortora:2021}, we conduct our analysis using the variables in their original scale.

Since annual expenditure cannot take negative values, variables used in the segmentation process are naturally bounded at zero. Standard clustering techniques often assume unbounded data, which is clearly not appropriate in the current context and, by failing to capture the true underlying structure, can lead to suboptimal segmentation.

Table~\ref{tbl:wholesale} reports the results obtained by fitting the standard GMM for unbounded variables, the GMMB model proposed in this paper, and some models discussed in \cite{Punzo:Tortora:2021}, namely the \emph{Mixture of Contaminated Normal} distributions (MCNM) and the \emph{Mixture of Multiple Scaled Contaminated Normal} distributions (MSCNM). For all models, the number of mixing components is fixed at $G=2$.
The MCNM model is fitted using the function \code{CNmixt()} in the \textbf{ContaminatedMixt} R package \citep{Punzo:Mazza:McNicholas:2018, Rpkg:ContaminatedMixt}, while the MSCNM is fitted using the function \code{mscn()} in the \textbf{MSclust} R package \citep{Rpkg:MSclust}. 
Models are compared using the BIC, ICL, NCE and the adjusted Rand index \citep[ARI;][]{Hubert:Arabie:1985}. 

Both the MCNM and MSCNM models provide significant improvements over the standard GMM, offering superior fit as evidenced by higher BIC values, and enhanced classification performance, with lower error rates (ER) and higher adjusted Rand index (ARI). However, both models are outperformed by the GMMB model, which achieves higher BIC and ICL values, indicating a better overall fit, while also approximately halving the ER and doubling the ARI, further emphasizing its superior classification capabilities.
As for the MCNM and MSCNM models, the classification uncertainty of GMMB, measured by NCE, is higher than that of GMM, reflecting a larger overlap among mixture components in GMMB model.

\begin{table*}[htb]
\caption{Model comparison for the clustering of the wholesale dataset, using Gaussian Mixture Model (GMM), Mixture of Contaminated Normals (MCNM), Mixture of Multiple Scaled Contaminated Normals (MSCNM), and Gaussian Mixture Model for Bounded data (GMMB), all fitted using $G=2$ mixture components.
Specification of models follow the \texttt{mclust} nomenclature \citep[see][Table 2.1]{mclust:book:2023}, so (VVV,2) denotes a model with two mixture components and unconstrained component-covariance matrices, while model VVE indicates heteroschedastic component-covariance matrices with equal orientation.}
\label{tbl:wholesale}
\centering
\begin{tabular}{@{}lrrrrrr@{}}
\toprule
Model  & log-likelihood & df &      BIC &        ICL &    NCE &   ARI \\
\midrule
GMM(VVV,2)  & -25069.70 & 55 & -50474.18 & -50495.52 & 0.0796 & 0.1028 \\
MCNM(VVV,2) & -24614.60 & 59 & -49588.32 & -49626.70 & 0.1439 & 0.3808 \\
MSCNM(2)    & -24498.57 & 79 & -49477.99 & -49491.15 & 0.1290 & 0.3642 \\
GMMB(VVE,2) & -23909.79 & 46 & -48099.57 & -48141.26 & 0.1539 & 0.6585 \\
\bottomrule
\end{tabular}
\end{table*}

\begin{figure}[htb]
\centering
\includegraphics[width=\textwidth]{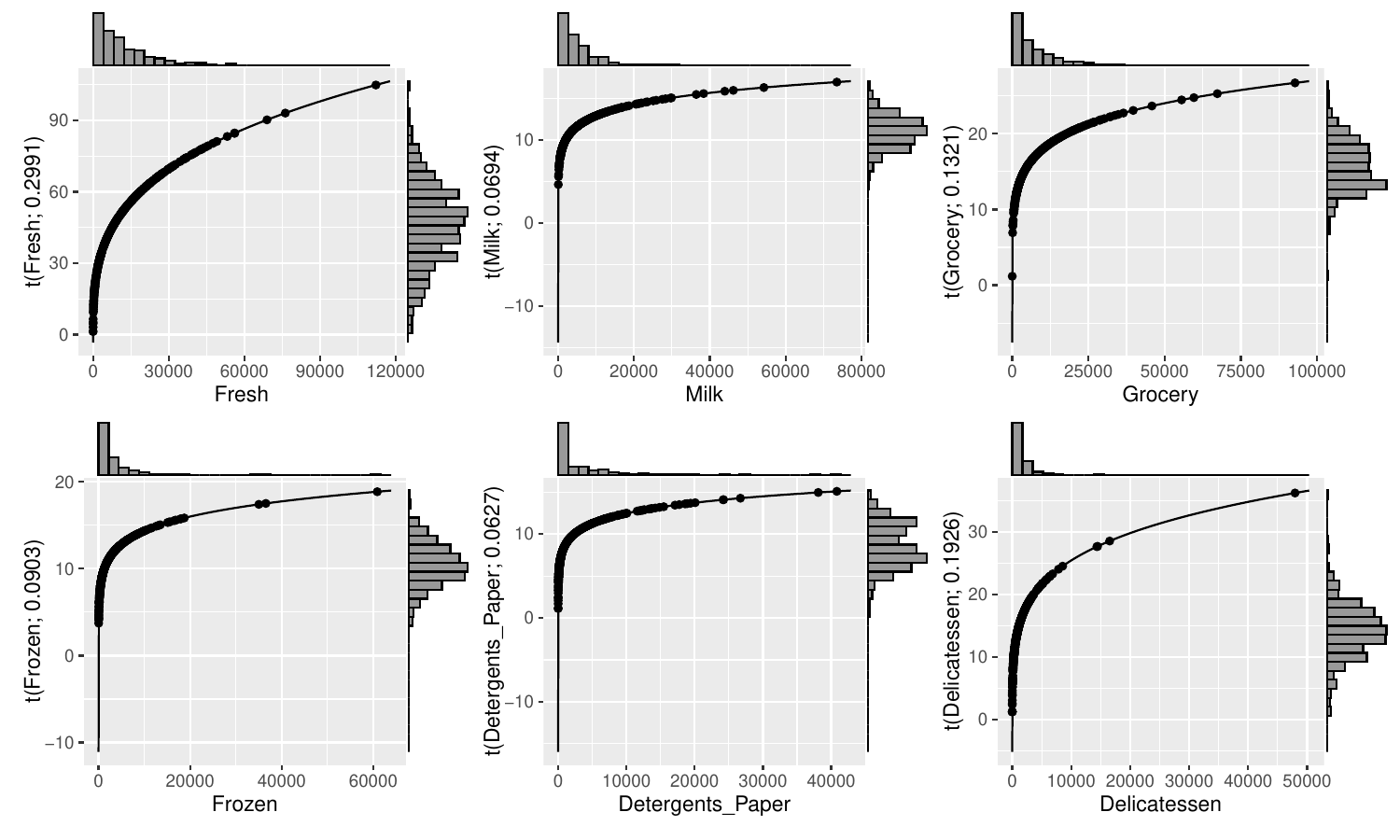}
\caption{Plots displaying the selected range-power transformations $t(X_j; \hat{\lambda})$ as a function of the original variables $X_j$ in the wholesale dataset, with marginal histograms showing the distribution before and after the applied transformation.}
\label{fig:wholesale1}
\end{figure}

Figure~\ref{fig:wholesale1} shows the range-power transformations in \eqref{eqn:range_power_transform_lb} for each variable in the wholesale dataset, as estimated by the GMMB model. 
All transformations exhibit a log-type shape, with $\lambda$ values ranging from $0.05$ to $0.3$. 
These transformations effectively reduce the pronounced positive skewness in the marginal distributions, as illustrated in the corresponding marginal histograms.

\begin{figure}[htb]
\centering
\includegraphics[width=\textwidth]{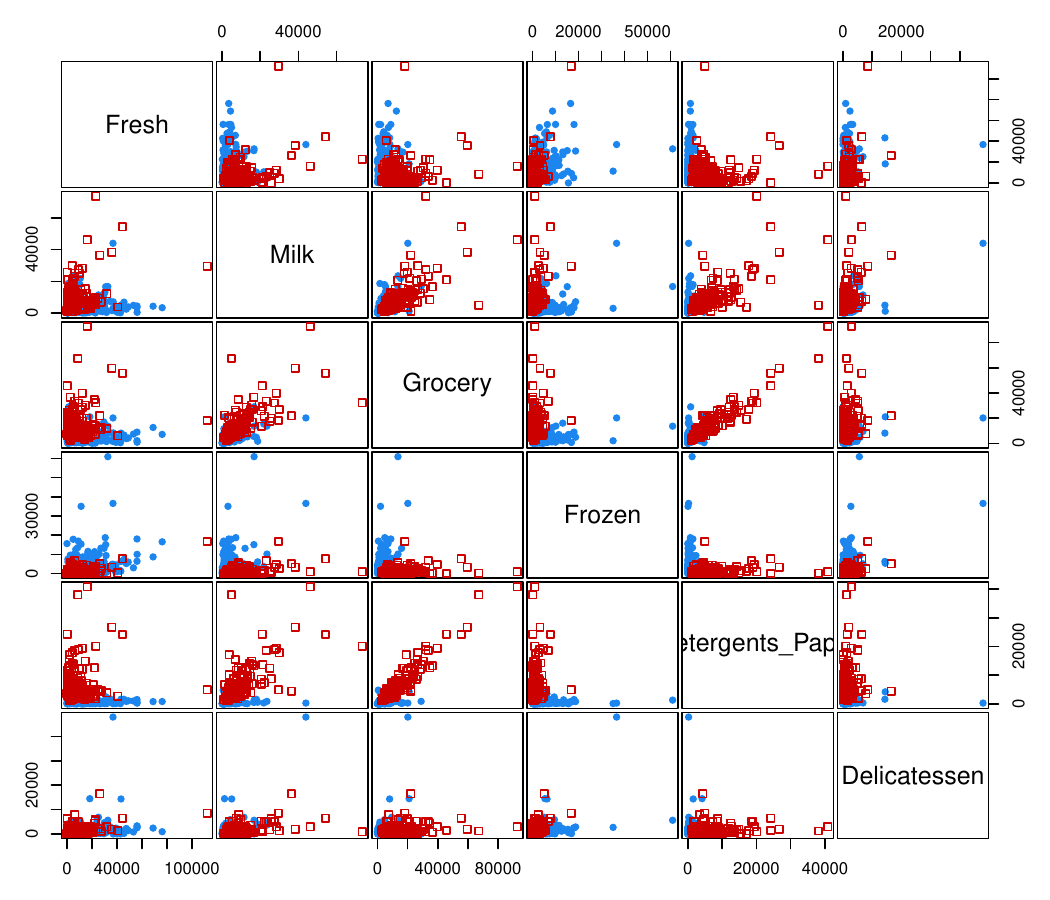}
\caption{Scatterplot matrix of variables in the wholesale dataset, with wholesale customers marked by their cluster membership.}
\label{fig:wholesale2}
\end{figure}

Figure~\ref{fig:wholesale2} presents the scatterplot matrix of the variables from the wholesale dataset, with points representing wholesale customers marked by their assigned cluster. The distributions within each segment exhibit noticeable skewness due to the zero-lower bound constraint of the variables, a characteristic effectively captured by the GMMB model used for the segmentation.

\begin{figure}[htb]
\centering
\includegraphics[width=0.8\textwidth]{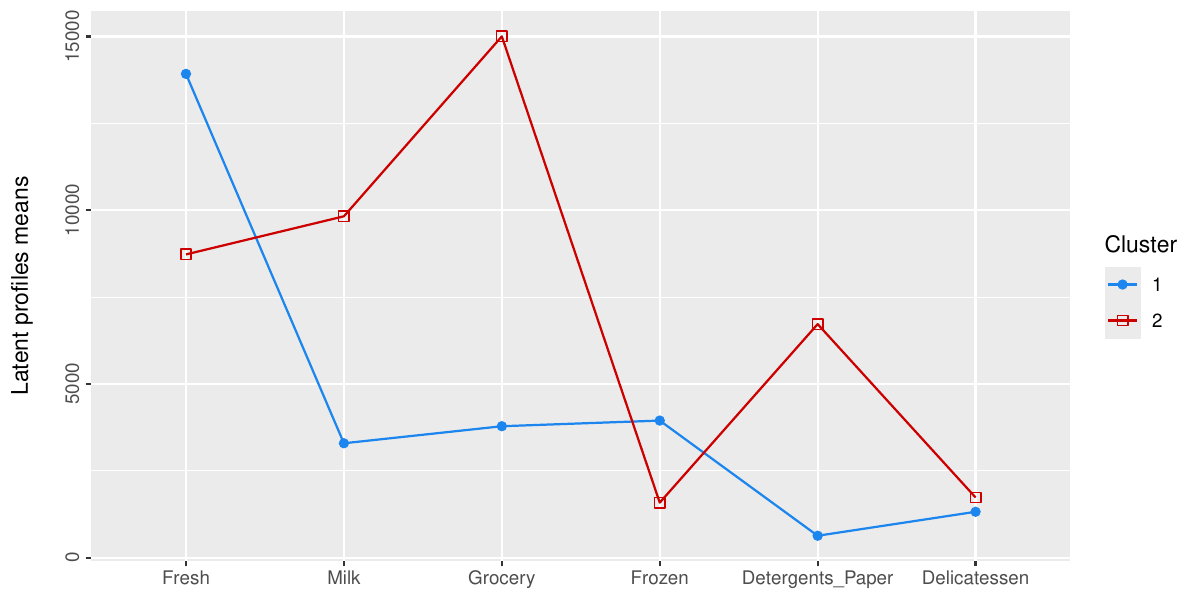}
\caption{Latent profiles plot showing the estimated cluster means from the fitted GMMB model for each variable in the wholesale dataset.}
\label{fig:wholesale3}
\end{figure}

Finally, Figure~\ref{fig:wholesale3} shows the average latent profiles for the two identified clusters. 
This graph clearly highlights the key variables that distinguish the two segments of wholesale distributor customers, with the ``delicatessen'' and ``frozen'' product categories not playing a significant role in the segmentation

\clearpage

\subsection{Human Development Index}
\label{sec:hdi}

The Human Development Index (HDI) is a composite index provided by the United Nations Development Programme (UNDP) and designed to measure the level of human development in world countries \citep{owid-hdi:2023}.

The HDI combines three key dimensions of human development: \emph{health}, measured by life expectancy at birth, \emph{education}, measured by expected years of schooling (for children of school entering age) and average years of schooling (for adults aged 25 and older), and \emph{standard of living}, measured by Gross National Income (GNI) per capita.
These three indicators are first normalized to have a minimum of 0 and a maximum of 1, then are combined by calculating the geometric mean.
The resulting HDI value for each country represents an overall score in the range $[0,1]$, with higher values indicating a better level of human development.

\begin{figure}[htb]
\centering
\includegraphics[width=\textwidth]{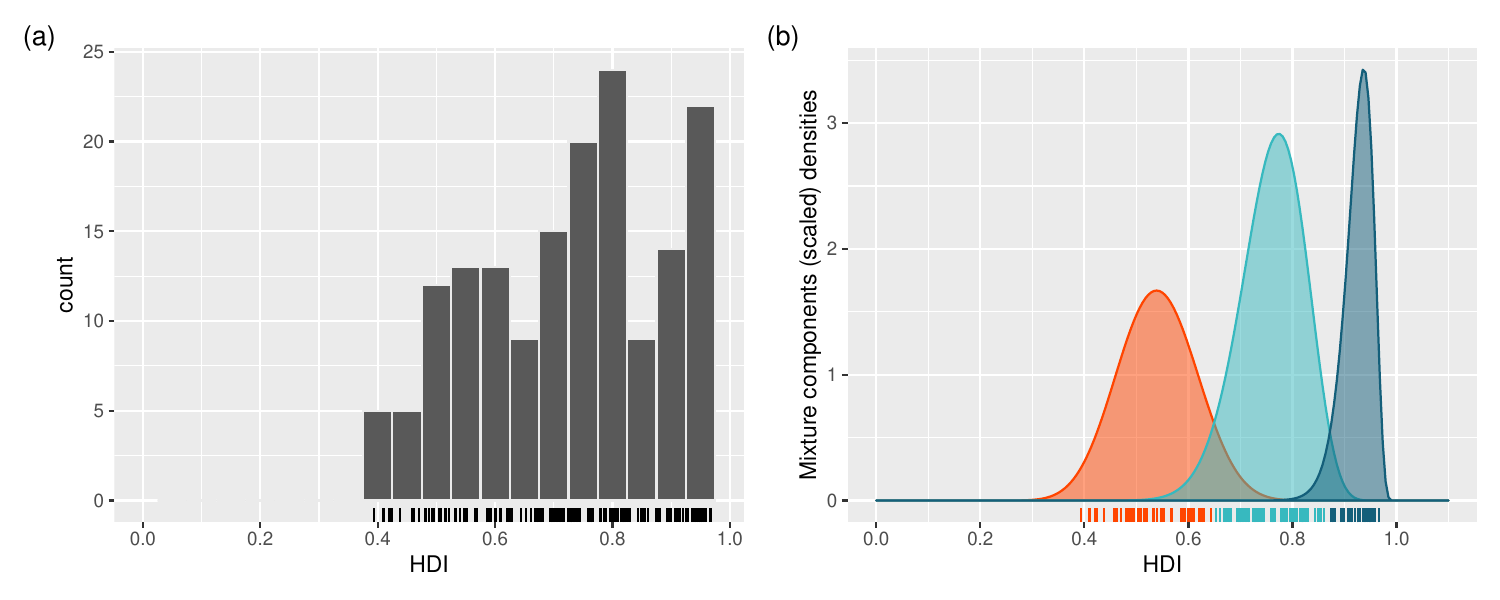}
\caption{(a) Histogram of the Human Development Index (HDI) distribution for the year 2022. (b) Plot of component densities, scaled by the corresponding mixing probabilities, estimated using the Gaussian Mixture Model for Bounded data (GMMB).}
\label{fig:hdi1}
\end{figure}

Figure~\ref{fig:hdi1}a shows the distribution of the Human Development Index (HDI) for the year 2022, revealing a clearly multimodal pattern. HDI values span a range from approximately 0.4 to just under 1, reflecting significant variation across countries.

According to the BIC criterion, the best-fitting GMMB model comprises three components with equal variance and a range-power transformation parameter in \eqref{eqn:range_power_transform_lb_ub} equals to $\hat{\lambda} = -0.12$. 
The corresponding component densities (scaled by their mixing probabilities) are shown in Figure~\ref{fig:hdi1}b. This model indicates the presence of three distinct clusters of countries, which can be broadly categorized as low, medium, and high human development clusters. 

\begin{figure}[htb]
\centering
\includegraphics[width=\textwidth]{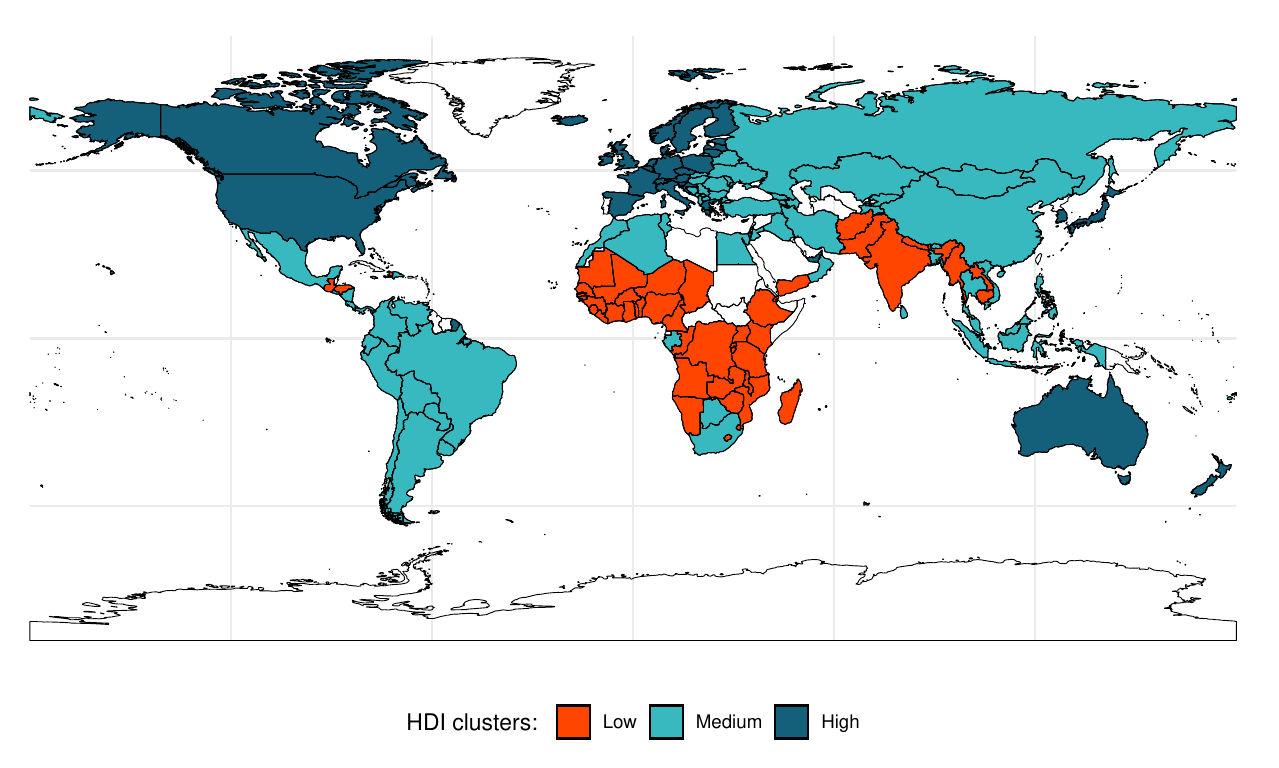}
\caption{World map showing countries clustered by Human Development Index (HDI) using the Gaussian Mixture Model for Bounded data (GMMB).}
\label{fig:hdi2}
\end{figure}

Figure~\ref{fig:hdi2} presents a world map with countries colored according to their respective cluster memberships.
The high-HDI cluster includes mainly developed countries, such as those in North America and Europe, along with Japan, Australia, New Zealand, and the United Arab Emirates.
The low-HDI cluster consists mostly of underdeveloped countries, primarily in Africa, with additional members in Asia (Afghanistan, Pakistan, Yemen, Nepal, Myanmar, Laos, Cambodia) and Central America (Guatemala, Haiti, Honduras). 
The remaining countries, mainly from Central and South America, Eastern Europe, and parts of Asia, fall into the medium-HDI cluster.

The selected GMMB model can be compared to the standard Gaussian Mixture Model (GMM) and a Mixture of Beta distributions \citep[BMM;][]{Bagnato:Punzo:2013, Dean:Nugent:2013}, with the number of components again selected using the BIC criterion. 
The results in Table~\ref{tbl:hdi} indicate that the GMMB model achieves a better fit, as evidenced by a higher BIC, along with a slightly reduced classification uncertainty, as shown by a lower NCE. 
Overall, the GMMB approach offers a reasonable and interpretable clustering solution with lowered uncertainty, further highlighting its effectiveness in this context.

\begin{table*}[htb]
\caption{Model comparison for the clustering of the Human Development Index (HDI), using Gaussian Mixture Model (GMM), Beta Mixture Model (BMM), and Gaussian Mixture Model for Bounded data (GMMB), with the number of components selected according to the BIC criterion.
Following \texttt{mclust} nomenclature \citep[see][Table 2.1]{mclust:book:2023}, model (V,3) indicates a model with three mixture components and varying variances, while (E,3) denotes a model with three mixture components and equal variance across components.}
\label{tbl:hdi}
\centering
\begin{tabular}{@{}lrrrrr@{}}
\toprule
Model     & log-likelihood & df & BIC & ICL & NCE \\
\midrule
GMM(V,3)  & 96.6144 & 8 & 152.5775 & 127.8651 & 0.1660 \\
BMM(3)    & 97.8045 & 8 & 154.9578 & 128.7258 & 0.1609 \\
GMMB(E,3) & 97.8727 & 7 & 160.1756 & 135.8538 & 0.1575 \\
\bottomrule
\end{tabular}
\end{table*}

\clearpage

\section{Conclusion}
\label{sec:conclusion}

This study aimed to develop a model-based clustering approach for bounded data, extending earlier work of \cite{Scrucca:2019} on Gaussian mixture density estimation. The primary research question focused on overcoming the unique challenges posed by bounded data, particularly by transforming bounded variables into an unbounded space to facilitate the application of GMMs.
Through the proposed range-power transformation, our method provides a flexible and interpretable clustering framework that preserves data bounds, enhancing the accuracy and applicability of GMMs for real-world bounded data scenarios.

The analyzed datasets demonstrated that this transformation-based approach offers a unified and flexible framework capable of handling diverse bounded data structures while ensuring model parsimony and interpretability. 
Compared to traditional GMMs and other model-based approaches using distributions with limited support, our method showed improved clustering partitions and interpretability, highlighting its relevance for practical applications involving the analysis of bounded data.

Despite the advantages mentioned above, some limitations remain. The proposed transformation-based approach requires selecting a set of appropriate global coordinate-wise transformation parameters. However, this approach may be suboptimal if different transformations are needed for different clusters or subsets of the data, a topic warranting further exploration in future research.

Additionally, the suitability of the proposal for handling heavy-tailed clusters should be investigated. This would allow to extend the transformation-based approach also in case of unbounded but non-spherical cluster distributions, hence offering an alternative approach to the several existing methods for non-Gaussian clusters \citep{McNicholas:2016}.


\bigskip

\paragraph{Code and Data Availability}
All the analyses have been conducted in R \citep{Rstat} using the \code{mclust} and \code{mclustAddons} packages \citep{Rpkg:mclust, Rpkg:mclustAddons}.
Source code and datasets to reproduce the analyses are available in a GitHub repository at 
\url{https://github.com/luca-scr/MclustBounded}. 

\bigskip


\clearpage
\bibliography{paper_arxiv}

\begin{thebibliography}{}

\bibitem[Bagnato and Punzo, 2013]{Bagnato:Punzo:2013}
Bagnato, L. and Punzo, A. (2013).
\newblock Finite mixtures of unimodal beta and gamma densities and the
  $k$-bumps algorithm.
\newblock {\em Computational Statistics}, 28(4):1571--1597.

\bibitem[Banfield and Raftery, 1993]{Banfield:Raftery:1993}
Banfield, J. and Raftery, A.~E. (1993).
\newblock Model-based {G}aussian and non-{G}aussian clustering.
\newblock {\em Biometrics}, 49:803--821.

\bibitem[Bechtel et~al., 1993]{Bechtel:etal:1993}
Bechtel, Y.~C., Bonaiti-Pellie, C., Poisson, N., Magnette, J., and Bechtel,
  P.~R. (1993).
\newblock A population and family study {N}-acetyltransferase using caffeine
  urinary metabolites.
\newblock {\em Clinical Pharmacology \& Therapeutics}, 54(2):134--141.

\bibitem[Biernacki et~al., 1999]{Biernacki:Celeux:Govaert:1999}
Biernacki, C., Celeux, G., and Govaert, G. (1999).
\newblock An improvement of the {NEC} criterion for assessing the number of
  clusters in a mixture model.
\newblock {\em Pattern Recognition Letters}, 20(3):267–--272.

\bibitem[Biernacki et~al., 2000]{Biernacki:Celeux:Govaert:2000}
Biernacki, C., Celeux, G., and Govaert, G. (2000).
\newblock Assessing a mixture model for clustering with the integrated
  completed likelihood.
\newblock {\em {IEEE} Transactions on Pattern Analysis and Machine
  Intelligence}, 22(7):719--725.

\bibitem[Bouveyron et~al., 2019]{Bouveyron:Celeux:Murphy:Raftery:2019}
Bouveyron, C., Celeux, G., Murphy, T.~B., and Raftery, A.~E. (2019).
\newblock {\em Model-Based Clustering and Classification for Data Science: With
  Applications in {R}}.
\newblock Cambridge University Press.

\bibitem[Browne and McNicholas, 2015]{Browne:McNicholas:2015}
Browne, R.~P. and McNicholas, P.~D. (2015).
\newblock A mixture of generalized hyperbolic distributions.
\newblock {\em Canadian Journal of Statistics}, 43(2):176--198.

\bibitem[Byrd et~al., 1995]{Byrd:etal:1995}
Byrd, R.~H., Lu, P., Nocedal, J., and Zhu, C. (1995).
\newblock A limited memory algorithm for bound constrained optimization.
\newblock {\em SIAM Journal on Scientific Computing}, 16(5):1190--1208.

\bibitem[Cardoso, 2013]{uci_wholesale}
Cardoso, M. (2013).
\newblock {Wholesale customers}.
\newblock UCI Machine Learning Repository.

\bibitem[Celeux and Govaert, 1995]{Celeux:Govaert:1995}
Celeux, G. and Govaert, G. (1995).
\newblock {G}aussian parsimonious clustering models.
\newblock {\em Pattern Recognition}, 28:781--793.

\bibitem[Celeux and Soromenho, 1996]{Celeux:Soromenho:1996}
Celeux, G. and Soromenho, G. (1996).
\newblock An entropy criterion for assessing the number of clusters in a
  mixture model.
\newblock {\em Journal of Classification}, 13(2):195--212.

\bibitem[Dean and Nugent, 2013]{Dean:Nugent:2013}
Dean, N. and Nugent, R. (2013).
\newblock Clustering student skill set profiles in a unit hypercube using
  mixtures of multivariate betas.
\newblock {\em Advances in Data Analysis and Classification}, 7(3):339--357.

\bibitem[Dempster et~al., 1977]{Dempster:Laird:Rubin:1977}
Dempster, A.~P., Laird, N.~M., and Rubin, D.~B. (1977).
\newblock Maximum likelihood from incomplete data via the {EM} algorithm (with
  discussion).
\newblock {\em Journal of the Royal Statistical Society: Series B (Statistical
  Methodology)}, 39:1--38.

\bibitem[Fraley and Raftery, 2002]{Fraley:Raftery:2002}
Fraley, C. and Raftery, A.~E. (2002).
\newblock Model-based clustering, discriminant analysis, and density
  estimation.
\newblock {\em Journal of the American Statistical Association},
  97(458):611--631.

\bibitem[Fraley et~al., 2024]{Rpkg:mclust}
Fraley, C., Raftery, A.~E., and Scrucca, L. (2024).
\newblock {\em mclust: Gaussian Mixture Modelling for Model-Based Clustering,
  Classification, and Density Estimation}.
\newblock {R} package version 6.1.1.

\bibitem[Gallaugher et~al., 2020]{Gallaugher:etal:2020}
Gallaugher, M. P.~B., McNicholas, P.~D., Melnykov, V., and Zhu, X. (2020).
\newblock Skewed distributions or transformations? modelling skewness for a
  cluster analysis.

\bibitem[Gormley et~al., 2023]{Gormley:Murphy:Raftery:2023}
Gormley, I.~C., Murphy, T.~B., and Raftery, A.~E. (2023).
\newblock Model-based clustering.
\newblock {\em Annual Review of Statistics and Its Application}, 10:573--595.

\bibitem[Hedelin and Skoglund, 2000]{Hedelin:Skoglund:2000}
Hedelin, P. and Skoglund, J. (2000).
\newblock Vector quantization based on {G}aussian mixture models.
\newblock {\em IEEE Transactions on Speech and Audio Processing},
  8(4):385--401.

\bibitem[Herre and Arriagada, 2023]{owid-hdi:2023}
Herre, B. and Arriagada, P. (2023).
\newblock The human development index and related indices: what they are and
  what we can learn from them.
\newblock {\em Our World in Data}.

\bibitem[Hubert and Arabie, 1985]{Hubert:Arabie:1985}
Hubert, L. and Arabie, P. (1985).
\newblock Comparing partitions.
\newblock {\em Journal of Classification}, 2:193--218.

\bibitem[John, 1970]{John:1970}
John, S. (1970).
\newblock On identifying the population of origin of each observation in a
  mixture of observations from two gamma populations.
\newblock {\em Technometrics}, 12(3):565--568.

\bibitem[Karlis and Santourian, 2009]{Karlis:Santourian:2009}
Karlis, D. and Santourian, A. (2009).
\newblock Model-based clustering with non-elliptically contoured distributions.
\newblock {\em Statistics and Computing}, 19(1):73--83.

\bibitem[Lindblom and Samuelsson, 2003]{Lindblom:Samuelsson:2003}
Lindblom, J. and Samuelsson, J. (2003).
\newblock Bounded support {G}aussian mixture modeling of speech spectra.
\newblock {\em IEEE Transactions on Speech and Audio Processing}, 11(1):88--99.

\bibitem[Lo and Gottardo, 2012]{Lo:Gottardo:2012}
Lo, K. and Gottardo, R. (2012).
\newblock Flexible mixture modeling via the multivariate t distribution with
  the box-cox transformation: an alternative to the skew-t distribution.
\newblock {\em Statistics and computing}, 22(1):33--52.

\bibitem[Manly, 1976]{Manly:1976}
Manly, B. F.~J. (1976).
\newblock Exponential data transformations.
\newblock {\em The Statistician}, 25(1):37--42.

\bibitem[McLachlan and Krishnan, 2008]{McLachlan:Krishnan:2008}
McLachlan, G.~J. and Krishnan, T. (2008).
\newblock {\em The {EM} Algorithm and Extensions}.
\newblock Wiley-Interscience, Hoboken, New Jersey, 2nd edition.

\bibitem[McLachlan et~al., 2019]{McLachlan:etal:2019}
McLachlan, G.~J., Lee, S.~X., and Rathnayake, S.~I. (2019).
\newblock Finite mixture models.
\newblock {\em Annual Review of Statistics and Its Application}, 6(1):355--378.

\bibitem[McLachlan and Peel, 2000]{McLachlan:Peel:2000}
McLachlan, G.~J. and Peel, D. (2000).
\newblock {\em Finite Mixture Models}.
\newblock Wiley, New York.

\bibitem[McNicholas, 2016]{McNicholas:2016}
McNicholas, P.~D. (2016).
\newblock {\em Mixture Model-Based Classification}.
\newblock Chapman \& Hall/CRC.

\bibitem[McNicholas et~al., 2017]{McNicholas:etal:2017}
McNicholas, S.~M., McNicholas, P.~D., and Browne, R.~P. (2017).
\newblock A mixture of variance-gamma factor analyzers.
\newblock In Ahmed, S.~E., editor, {\em Big and Complex Data Analysis:
  Methodologies and Applications}, pages 369--385, Cham. Springer International
  Publishing.

\bibitem[Meng and Rubin, 1993]{Meng:Rubin:1993}
Meng, X.-L. and Rubin, D.~B. (1993).
\newblock Maximum likelihood estimation via the {ECM} algorithm: A general
  framework.
\newblock {\em Biometrika}, 80(2):267--278.

\bibitem[Punzo et~al., 2018]{Punzo:Mazza:McNicholas:2018}
Punzo, A., Mazza, A., and McNicholas, P.~D. (2018).
\newblock {ContaminatedMixt}: An {R} package for fitting parsimonious mixtures
  of multivariate contaminated normal distributions.
\newblock {\em Journal of Statistical Software}, 85(10):1--25.

\bibitem[Punzo et~al., 2023]{Rpkg:ContaminatedMixt}
Punzo, A., Mazza, A., and McNicholas, P.~D. (2023).
\newblock {\em {ContaminatedMixt}: Clustering and Classification with the
  Contaminated Normal}.
\newblock R package version 1.3.8.

\bibitem[Punzo and Tortora, 2021]{Punzo:Tortora:2021}
Punzo, A. and Tortora, C. (2021).
\newblock Multiple scaled contaminated normal distribution and its application
  in clustering.
\newblock {\em Statistical Modelling}, 21(4):332--358.

\bibitem[{R Core Team}, 2025]{Rstat}
{R Core Team} (2025).
\newblock {\em R: A Language and Environment for Statistical Computing}.
\newblock R Foundation for Statistical Computing, Vienna, Austria.

\bibitem[Richardson and Green, 1997]{Richardson:Green:1997}
Richardson, S. and Green, P.~J. (1997).
\newblock On bayesian analysis of mixtures with an unknown number of components
  (with discussion).
\newblock {\em Journal of the Royal Statistical Society: Series B (Statistical
  Methodology)}, 59(4):731--792.

\bibitem[Schwarz, 1978]{Schwarz:1978}
Schwarz, G. (1978).
\newblock Estimating the dimension of a model.
\newblock {\em The Annals of Statistics}, 6(2):461--464.

\bibitem[Scrucca, 2019]{Scrucca:2019}
Scrucca, L. (2019).
\newblock A transformation-based approach to {G}aussian mixture density
  estimation for bounded data.
\newblock {\em Biometrical Journal}, 61(4):873--888.

\bibitem[Scrucca, 2025]{Rpkg:mclustAddons}
Scrucca, L. (2025).
\newblock {\em {mclustAddons}: Addons for the 'mclust' Package}.
\newblock {R} package version 0.9.2.

\bibitem[Scrucca et~al., 2023]{mclust:book:2023}
Scrucca, L., Fraley, C., Murphy, T.~B., and Raftery, A.~E. (2023).
\newblock {\em Model-Based Clustering, Classification, and Density Estimation
  Using {mclust} in {R}}.
\newblock Chapman \& Hall/CRC, Boca Raton, FL.

\bibitem[Scrucca and Raftery, 2015]{Scrucca:Raftery:2015}
Scrucca, L. and Raftery, A.~E. (2015).
\newblock Improved initialisation of model-based clustering using {G}aussian
  hierarchical partitions.
\newblock {\em Advances in Data Analysis and Classification}, 4(9):447--460.

\bibitem[Tong and Tortora, 2024]{Rpkg:MixtureMissing}
Tong, H. and Tortora, C. (2024).
\newblock {\em {MixtureMissing}: Robust and Flexible Model-Based Clustering for
  Data Sets with Missing Values at Random}.
\newblock R package version 3.0.2.

\bibitem[Tortora et~al., 2024]{Rpkg:MSclust}
Tortora, C., Punzo, A., and Tran, L. (2024).
\newblock {\em {MSclust}: Multiple-Scaled Clustering}.
\newblock R package version 1.0.4.

\bibitem[Wiper et~al., 2001]{Wiper:Insua:Ruggeri:2001}
Wiper, M., Insua, D.~R., and Ruggeri, F. (2001).
\newblock Mixtures of gamma distributions with applications.
\newblock {\em Journal of Computational and Graphical Statistics},
  10(3):440--454.

\bibitem[Yeo and Johnson, 2000]{Yeo:Johnson:2000}
Yeo, I.-K. and Johnson, R.~A. (2000).
\newblock A new family of power transformations to improve normality or
  symmetry.
\newblock {\em Biometrika}, 87(4):954--959.

\bibitem[Young et~al., 2019]{Young:etal:2019}
Young, D.~S., Chen, X., Hewage, D.~C., and Nilo-Poyanco, R. (2019).
\newblock Finite mixture-of-gamma distributions: estimation, inference, and
  model-based clustering.
\newblock {\em Advances in Data Analysis and Classification}, 13(4):1053--1082.

\bibitem[Zhu and Melnykov, 2018]{Zhu:Melnykov:2018}
Zhu, X. and Melnykov, V. (2018).
\newblock Manly transformation in finite mixture modeling.
\newblock {\em Computational Statistics \& Data Analysis}, 121:190--208.

\end{thebibliography}

\end{document}